\documentclass[%
reprint,
nofootinbib,
amsmath,amssymb,
aps,
]{revtex4-1}

\usepackage[export]{adjustbox}
\usepackage{color}
\usepackage{graphicx}
\usepackage{dcolumn}
\usepackage{bm}
\usepackage{hyperref}
\usepackage{float}
\usepackage{subfigure}
\usepackage{hyperref}
\hypersetup{
    colorlinks=true,
    linktoc=page,    
    linkcolor=blue,
    urlcolor=magenta
}

\newcommand{\be}{\begin{equation}}
\newcommand{\ee}{\end{equation}}
\newcommand{\bea}{\begin{eqnarray}}
\newcommand{\eea}{\end{eqnarray}}

\newcommand{\ket}[1]{\left|{#1}\right>}

\begin{document}

\title{Circumventing the black hole hair-loss problem}

\author{Peng Cheng$^{1,2}$}
\email{p.cheng.nl@outlook.com}

\affiliation{1) Center for Joint Quantum Studies and Department of Physics, 
Tianjin University, Tianjin 300350, China\\
2) Lanzhou Center for Theoretical Physics, Key Laboratory of Theoretical Physics of Gansu Province, and Key Laboratory of Quantum Theory and Applications of MoE, Lanzhou University, Lanzhou, Gansu 730000, China
}

\begin{abstract}
We provide a possible way of avoiding the hair-loss problem by studying gauge fluctuations on classical Schwarzschild black hole background. The hair-loss problem arises due to the small fidelity of reconstructing the interior operator at the end stage of the evaporation, which is general in most schemes trying to decouple the early and late radiation to avoid firewall. To circumvent the problem, we put the black hole in a cavity as a toy model, and study the entropy behavior of the system with different temperatures. By analyzing gauge fluctuations with nontrivial boundary conditions, we find that the entropy of the black hole system increases at the end stage, rather than directly dropping to zero. Besides the end stage, the entropy is the same as Page's original argument. The hair-loss problem can be avoided by the proposed model and we can gain important insights into the end stage of the evaporation and quantum effects of gravity.
\end{abstract}

\maketitle
\flushbottom



\section{Introduction}
\label{intro}

The black hole information paradox (BHIP) is one of the most important subjects in black hole physics, which has brought us tons of new insights about the quantum properties of gravity and holography. The paradox was put forward by Hawking \cite{Hawking:1976ra}, and has burst with new vitality again and again in the past half century. It is worthwhile mentioning that the recent breakthrough related to the island rule \cite{Penington2020, Almheiri:2019psf, Almheiri:2019hni} has renewed our understanding of Euclidean wormholes \cite{Penington:2019kki, Almheiri:2019qdq}, holography \cite{Marolf:2020xie, Chen2020, Chen2020a, Grimaldi2022, Hernandez:2020nem, Geng:2020qvw, Geng:2021hlu}, and more \cite{Marolf:2020rpm, Marolf:2021ghr}.

To gain a microscopic understanding of the information paradox, one always needs to address the problem from an information-theoretical approach, and the Almheiri-Marolf-Polchinski-Sully 
 (AMPS) firewall \cite{Almheiri:2012rt} is the key to the whole story. The Page curve \cite{Page:1993wv, Page:2013dx} of the radiation entropy consistent with the unitarity implies the late-time Hawking radiation $H$ and its partner $P$ cannot be entangled with each other after the Page time due to the monogamy of entanglement. The firewall arises because the near horizon region is not equivalent to the Minkowski vacuum. To avoid firewall, one needs to decouple the early and late radiation, for example the Harlow-Hayden-Aaronson (HHA) decoding task \cite{Harlow:2013tf, Aaronson:2016vto}.

To achieve a more explicit decoupling between the early and late radiation and reconstruct the interior operator, recent studies suggest that we need to use soft hair degrees of freedom.
Black hole soft hair was used to understand black hole microstates and suggested to solve the information paradox since it was put forward \cite{Strominger2020, Hawking:2016msc, Hawking:2016sgy}.
Significative improvement was made by Yoshida's decoupling theorem \cite{Yoshida:2018ybz, Yoshida:2019kyp, Yoshida:2017non}, where he proved the possibility of decoupling the early and late radiation in the presence of soft hair. Inspired by that, recent progress has demonstrated an explicit realization of reconstruction of the interior operator using the soft hair degrees of freedom \cite{Yoshida:2018ybz, Yoshida:2019kyp, Yoshida:2017non, Pasterski:2020xvn, Cheng:2020vzw}. 
 
However, the soft BHIP program is not perfect, and the hair-loss problem \cite{Cheng:2021gdr} is a conspicuous example. 
The decoupling between the early and late radiation and reconstruction of interior operators heavily rely on a large phase space of soft hair. Moreover, the related fidelity is inversely proportional to the size of the soft hair squared. The so-called hair-loss problem is about the fact that the black hole cannot support a large soft hair phase space at the end of the evaporation and the soft BHIP program falls down. At the end of the evaporation, small coarse-grained entropy cannot support any scheme decoupling the early and late radiation. So the hair-loss problem is general in schemes like the HHA scheme.

This paper tries to address the hair-loss problem and provide a possible solution to the problem by studying a toy model for gravitational fluctuation on black holes. The toy model is a gauge theory with non-trivial boundary conditions. We can perform Euclidean path integral to evaluate the partition function and corresponding entropy. The final result suggests that for an evaporating black hole, the beginning of the evaporation was not modified. While at the end stage of the evaporation the entropy does not drop to zero and the soft BHIP argument is saved.

\section{The hair-loss problem}
\label{hair-loss}

Soft hair plays an important role in understanding the BHIP.  
Although important breakthroughs have been made, the soft BHIP program is not perfect.
In this section, we briefly discuss the hair-loss problem and related issues.

\begin{figure}
\centering
\includegraphics[width=0.4\textwidth]{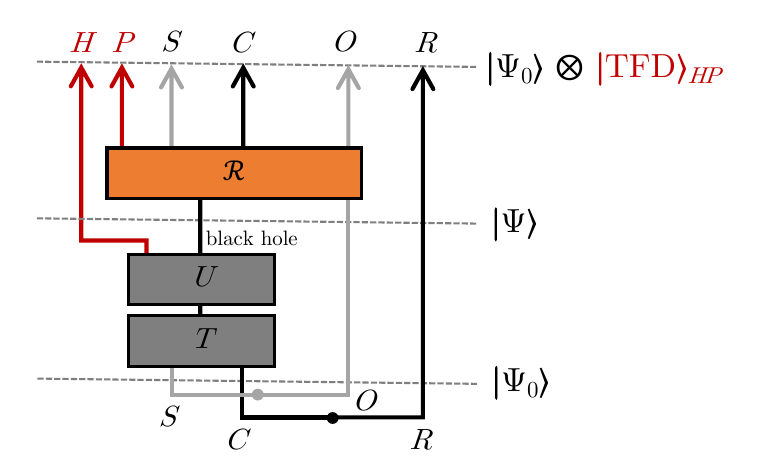}  \caption{After the Page time, for a system $\ket{\Psi_0}$ contains code subspace of a black hole $C$, early radiation $R$, soft hair $S$, and exterior observer $O$, there is an operation $T$ that map $C$ and $S$ to the black hole. After a unitary evolution $U$, an extra Hawking radiation $H$ is emitted. We are looking for a recovery operation $\mathcal{R}$ that can recover the original system $\ket{\Psi_0}$ and a thermo-field double state composing Hawking radiation $H$ and its partner $P$. The recovery process should not use the early radiation degrees of freedom $R$.}\label{pic1}
\end{figure}

Let us look at the situation shown in Fig. \ref{pic1}, where we have shown a post-Page time radiation and a recovery process. The recovery process $\mathcal{R}$ is able to reconstruct the partner $P$ of late-time Hawking radiation $H$ without using any information of early radiation $R$. This means that the early and late radiations are decoupled with each other in the presence of soft hair degrees of freedom. The decoupling between early and late radiation dissolves the firewall, so we naturally get a Page curve consistent with the unitary evolution of the black hole. It was shown that the recovery process can be accomplished with the help of the Petz map \cite{Pasterski:2020xvn}.

The fidelity of the reconstruction can be evaluated using the almost perfect scrambling property of the black hole evolution \cite{Yoshida:2017non, Hosur:2015ylk, Cheng:2021gdr}. 
The error of the reconstruction process is of order ${d_H^2}/{d_S^2}$, i.e.
\be\label{error}
\text{Error}\sim \mathcal{O}\left(\frac{d_H^2}{d_S^2}\right)\,,
\ee
where $d_H$ and $d_S$ are the dimensions of Hilbert spaces of late-time Hawking radiation $H$ and soft hair $S$. The fidelity is the inverse ratio of the error. To have a trustable reconstruction, the error should always be much smaller than 1. This requirement cannot be fulfilled at the end of the evaporation and the firewall would come back again. So we have the hair-loss problem \cite{Cheng:2021gdr},which can be summarized as follows:

Even though the decoupling between the early and late radiation can be realized after the Page time, the information paradox is not solved. At the end stage of the black hole evaporation, the black hole system does not have large enough coarse-grained entropy to support the schemes to construct the interior operator and decouple the early and late radiation.

The coarse-grained entropy of an evaporating black hole is shown in Fig. \ref{BH2} a), where we can see that we would finally lose the tool to decouple the early and late radiation at the end of the evaporation.

The hair-loss problem is general in most schemes trying to reconstruct the interior operators using quantum information protocol. The reconstruction interior operator and decoupling between early and late radiations usually rely on a large phase space the observer can access to and would not work at the end of the evaporation. For example, the HHA decoding task is not super hard if the black hole system is left with a few qubits. 

\section{Circumventing the hair-loss problem}
\label{therapy}

In this section, we discuss a possible way to solve the hair-loss problem by carefully studying the entropy of the black hole system with boundary would-be gauge degrees of freedom. 

\subsection{U(1) gauge theory on black hole background}

In two of our previous papers \cite{Cheng:2023bcv, Cheng:2023cms}, we studied U(1) gauge theory as a toy model of gravitational fluctuations of Schwarzschild black holes at finite temperature. 
We have put two boundaries on the black hole background, one at the stretched horizon $r=r_s+\varepsilon$ and the other at distance $L$ from the horizon. $r_s$ is the radius of the Schwarzschild black hole, and $\varepsilon$ is a small distance cut-off. By carefully studying the thermal entropy of the system using the Euclidean path integral, we can find the behavior of the entropy at the end of the evaporation.

Asymptotic Anti-de Sitter (AdS) black hole is like putting the black hole inside of a box \cite{Wald1980, Ishibashi:2003jd, Ishibashi:2004wx}. The ratio of two parameters, the AdS radius $l_{AdS}$ and the inverse temperature $\beta$, can vary depending on the temperature of the system. For fixed $l_{AdS}$, $\beta$ can be large for low temperatures and small for high temperatures. 
Like the asymptotic AdS black hole, in this paper we use a toy model to mimic the setup and put two boundaries on evaporating black hole background. 
For super-low temperature (large) black holes, the inverse temperature $\beta$ can be superlarge. 
In such case, we have $r_s\gg L$, due to the relation between $\beta$ and $r_s$\footnote{We actually compare the inverse temperature $\beta$ with $L$. It is particularly useful for Reissner-Nordstr\"{o}m black hole, where $r_s$ and $\beta$ are not `identified', as discussed in \cite{Cheng:2023cms}. In that case, you cannot simply imagine we have a large or small size black hole in a cavity.
}.
Note that unlike $l_{AdS}$, $L$ is the distance away from the horizon, the situation when $r_s\gg L$ means that the radius of the cavity is close to the radius of the horizon.
For high temperatures, the inverse temperature is relatively small and correspondingly we have small black holes. 
$L$ can be much larger than $r_s$ in the case.
This can be regarded as the end stage of the evaporation of the Schwarzschild black hole, where the black hole is small and hot.

Defining $\rho=r-r_s$, we can write the Euclidean metric as
\be
ds^2=\frac{\rho}{\rho+r_s}d\tau^2+\frac{\rho+r_s}{\rho}d\rho^2+(\rho+r_s)^2d\Omega^2\,.
\ee
Now the boundaries are located at $\rho=\varepsilon$ and $\rho=L$. We can also define the proper distance away from the horizon as
\be
y=\sqrt{\rho(\rho+r_s)}+r_s~\text{arcsinh} \sqrt{\frac{\rho}{r_s}}\,,
\ee
and rewrite the metric in terms of $y$.
The Euclidean action for the U(1) gauge theory can be expressed as
\be
S_E=\frac{1}{4e^2}\int_{\mathcal{M}}d^4x\sqrt{g}~F^{\mu\nu}F_{\mu\nu}\,.
\ee
For boundaries with normal vector $n^{\mu}\partial_{\mu}=\partial_r$, the on-shell variation of the action is
\be
\delta S=-\frac{1}{e^2}\int_{\partial\mathcal{M}}d\tau d^2x\sqrt{h}~F^{ra}\delta A_a\,,
\ee
with induced metric $h_{ab}$ and $x^a=(\tau,\theta,\varphi)$. Without additional boundary terms, we have two obvious choices
\be\label{bcbc}
F^{ra}\Big{|}_{\partial\mathcal{M}}=0\,~~~\text{or}~~~\delta A_a\Big{|}_{\partial\mathcal{M}}=0\,.
\ee
Note that the other components $n^*_{\mu}F^{\mu r}\big{|}_{\partial\mathcal{M}}$ and $\delta A_a\big{|}_{\partial\mathcal{M}}$ are not constrained.
The Casimir boundary condition \cite{Milton2001, Bordag2009, Jaffe2005, Chernodub:2017gwe, Chernodub:2017mhi, Chernodub:2022izt} implies $n^*_{\mu}F^{\mu r}\big{|}_{\partial\mathcal{M}}=0$. So if we further ask $F^{ra}\big{|}_{\partial\mathcal{M}}$ in \eqref{bcbc} to be zero, the boundary would be too boring. Thus, the only interesting boundary condition in this setup is setting $\delta A_a\big{|}_{\partial\mathcal{M}}=0$, while letting $\delta A_a\big{|}_{\partial\mathcal{M}}$ to be arbitrary.
i.e. the following boundary conditions
\be
\delta A_a\Big{|}_{\partial \mathcal{M}}=0\,,~~~~\delta A_r\Big{|}_{\partial \mathcal{M}}=f(x^a)\,,\label{BC}
\ee
where $f(x^a)$ can have arbitrary dependence of $x^a$. It can be easily checked that the above boundary conditions give us a well-defined variation principle without adding any Gibbons-Hawking-like terms\footnote{See \cite{Cheng:2023bcv, Cheng:2023cms} for more physical reasons of the boundary condition.}. 

The gauge field $A_\mu$ can be separated into three parts: the bulk fluctuation modes $\hat{A}_\mu(x^\mu)$ that completely vanish at boundaries, zero longitudinal momentum modes of $A_r$ denoted as $\phi(x^a)$, and Wilson lines stretched between two boundaries $W(x^a)$. $\phi$ is the boundary modes on the stretched horizon, and $W$ defined as $i\int_{y_1}^{y_2} dy ~\hat{A}_y$ more or less captures the difference between the boundaries. ${y_1}$ and ${y_2}$ are locations of the two boundaries.  Note that $\phi$ and $W$ arise due to the boundary condition \eqref{BC}. The Euclidean path integral can be worked out after we get the effective action for $\phi$ and $W$ \cite{Cheng:2023cms}. The contribution from the bulk fluctuation modes can be expressed as
\be
Z_{\hat{A}}=\int \mathcal{D}\hat{A}_\mu\exp \left[-\frac{1}{4e^2}\int_{\mathcal{M}}d\tau d^3x~\sqrt{g}~\hat{F}^{\mu\nu}\hat{F}_{\mu\nu}\right].
\ee
Correspondingly, the entropy from those modes is
\bea
\mathcal{S}_{\hat A} && =\frac{16\pi^3}{45}\frac{1}{\beta^3}\frac{r_s^4}{\varepsilon}+\frac{64\pi^3}{45}\frac{r_s^3}{\beta^3}\ln\frac{L}{\varepsilon}\nonumber \\
&&~~~ +\frac{16\pi^3}{45}\frac{1}{\beta^3}\left(-\frac{r_s^4}{L}+6r_s^2L+3r_sL^2+\frac{L^3}{3}\right)\,.\label{7}
\eea
This contribution is similar to entropy derived from the brick wall model \cite{tHooft:1984kcu}.
Moreover, the partition function for modes $\phi$ and $W$ can be written as
\be
Z_{\phi,W}=Z_{\phi}\times Z_{W}=\int\mathcal{D}\phi ~\exp[-S_{\phi}]\times \int\mathcal{D}W ~\exp[-S_{W}]\,.
\ee
Let's take the effective action of $\phi$ as an example to illustrate the quantitive behavior of $\phi$ and $W$. 
In high-temperature limit $r_s\ll L$, the effective action can be approximated as
\bea\label{highS}
&& S_{\phi} =\frac{1}{2e^2L^2}\int d\tau d^2x ~L^2\sin\theta~ 
\\&&~~~~\times 
\Big(\frac{1}{3}\partial_\tau\phi\partial_\tau\phi + \frac{1}{L^2}\partial_\theta\phi\partial_\theta\phi+ \frac{1}{L^2\sin^2\theta}\partial_\varphi\phi\partial_\varphi\phi \Big)\,,\nonumber
\eea
the corresponding entropy can be directly evaluated as
\be
\mathcal{S}_{\phi}=(1-\beta\partial_\beta)\ln Z_{\phi}= \frac{\pi^3}{45}\frac{L^2}{\beta^2}\,.\label{10}
\ee
In the high-temperature limit, the entropy captures the fluctuation modes of a three-dimensional scalar field $\phi$ living in a manifold with topology $S^1\times S^2$. The length scales of $S^1$ and $ S^2$ are $\beta$ and $L$, separately.

In low-temperature limit $r_s\gg L$, the effective action can be written as
\bea
&& S_{\phi} =\frac{1}{2e^2(L+r_s)}\int d\tau d^2x ~r_s^2\sin\theta~ 
\nonumber\\&&~~~~~\times 
\Big(\frac{r_s}{L}\ln \frac{L}{\varepsilon}\partial_\tau\phi\partial_\tau\phi + \frac{1}{r_s^2}\partial_\theta\phi\partial_\theta\phi+ \frac{1}{r_s^2\sin^2\theta}\partial_\varphi\phi\partial_\varphi\phi \Big)\,.\nonumber\\ \label{lowS}
\eea
We would finally take the $\varepsilon\to 0$ limit, which introduces a localization on the space of zero energy modes $\partial_\tau\phi=0$ in the path integral. Because of the localization, the zero-point energy starts to contribute to the entropy, and the corresponding entropy is proportional to the area of the spatial space and UV cutoff. Supposing the cutoff is at the Planck scale, the entropy can be written as
\be\label{S1}
\mathcal{S}_\phi\propto\frac{r_s^2}{l_p^2}\,.
\ee
As can be seen from \eqref{7} and \eqref{10}, the entropy have different behaviors at low and high temperature. The change between different entropy behaviors will be useful for solving the hair-loss problem.

Note that the reason for the transition between the effective actions \eqref{highS} and \eqref{lowS} is that the coefficient in the action is proportional to a function
\be
F = 3r_s^2L+\frac{3}{2}r_s L^2+\frac{L^3}{3}+r_s^3\ln{L}/{\varepsilon}\,.\label{F}
\ee
$F$ can be approximated by $L^3/3$ when $L\gg r_s$, and $F$ is more or less $r^3_s\ln L/\varepsilon$ in the low temperature limit. 
 Similar behavior can also be demonstrated for the field $W$, whose effective action read as
{\small\bea
&&S_W=\frac{2}{e^2(L+r_s)}\int d\tau d^2x~r_s^2\sin\theta \left(\frac{r_s}{\varepsilon}\frac{Lr_s^2+r_s^3}{F}\partial_\tau W\partial_\tau W \right.\nonumber\\
&&~~~~~~~~\left.+\frac{1}{r_s^2}\partial_{\theta}W\partial_{\theta}W+\frac{1}{r_s^2\sin^2\theta}\partial_{\varphi}W\partial_{\varphi}W\right)\,.
\eea
}
It can be seen that the effective action $S_W$ has a similar behavior as demonstrated for field $\phi$. 
It is important to notice that the order of taking different limits is vital for the story.

\begin{figure}
\centering
\includegraphics[width=0.42\textwidth]{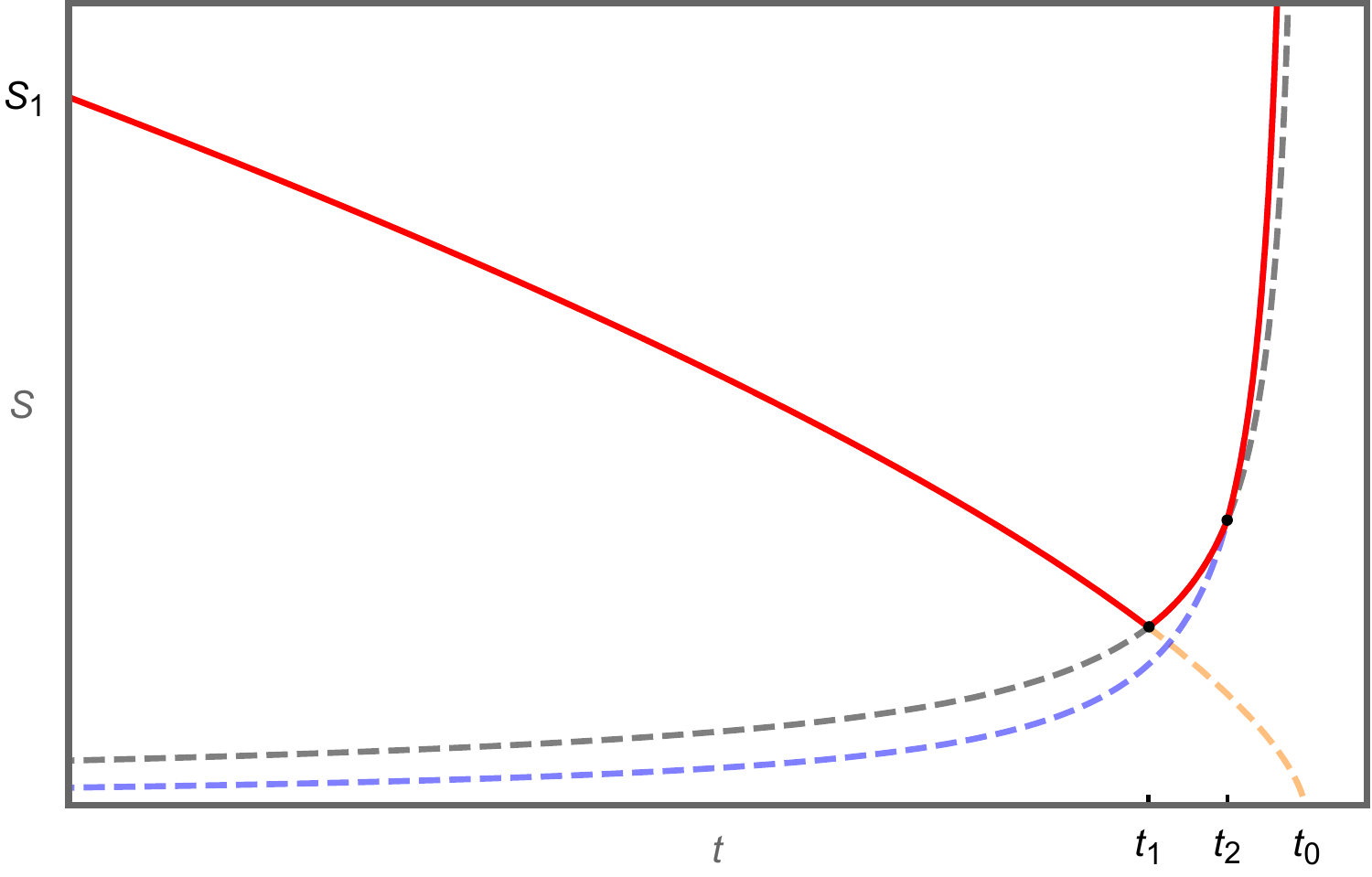}\\
\caption{
Sketch of different entropy behaviors for evaporating black holes. The overall evaporating time of the black hole is denoted as $t_0$. Before $t_1$, the entropy is proportional to $r_s^2$ as indicated in \eqref{S1}. After $t_1$, the contribution that scales as \eqref{10} starts to dominate. We get volume-dominated contribution after $t_2$.
}
\label{BH}
~\\~\\
\setcounter {subfigure} {0} a){\includegraphics[width=0.42\textwidth]{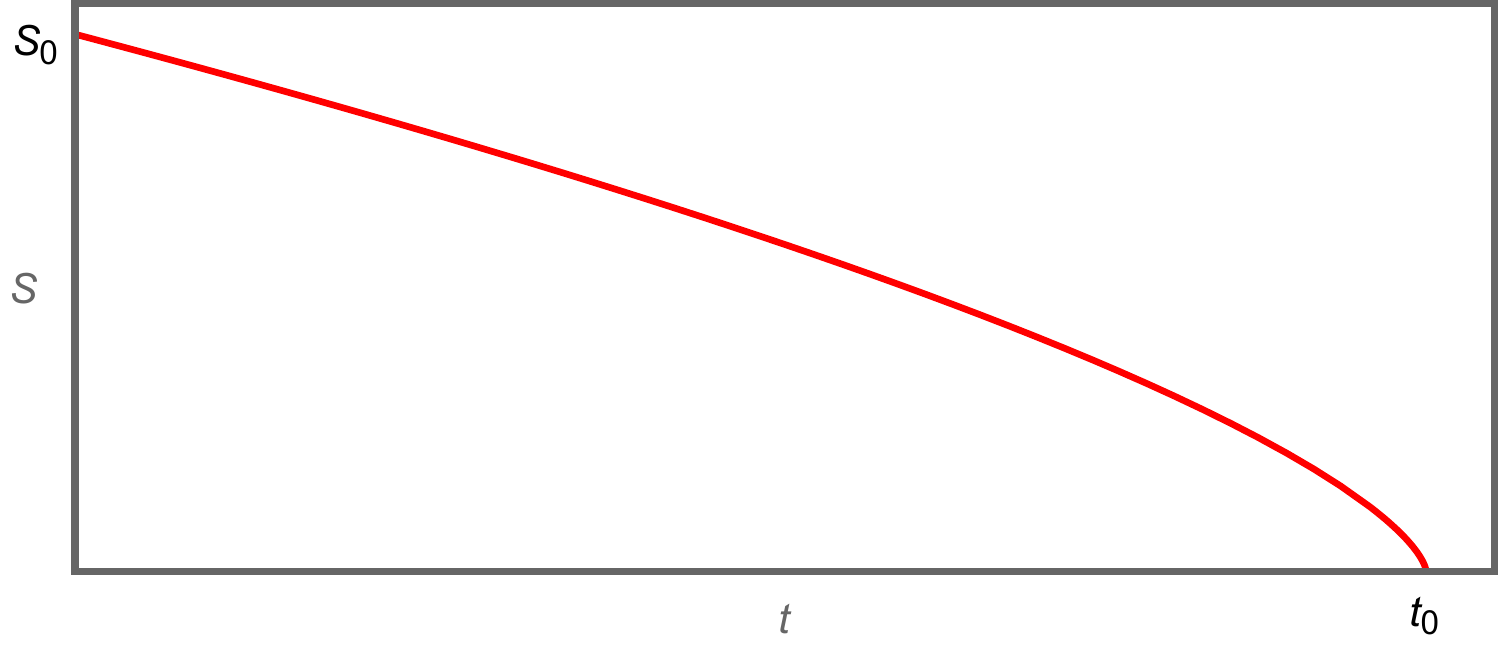}
}\\
\setcounter {subfigure} {0} b){
\includegraphics[width=0.42\textwidth]{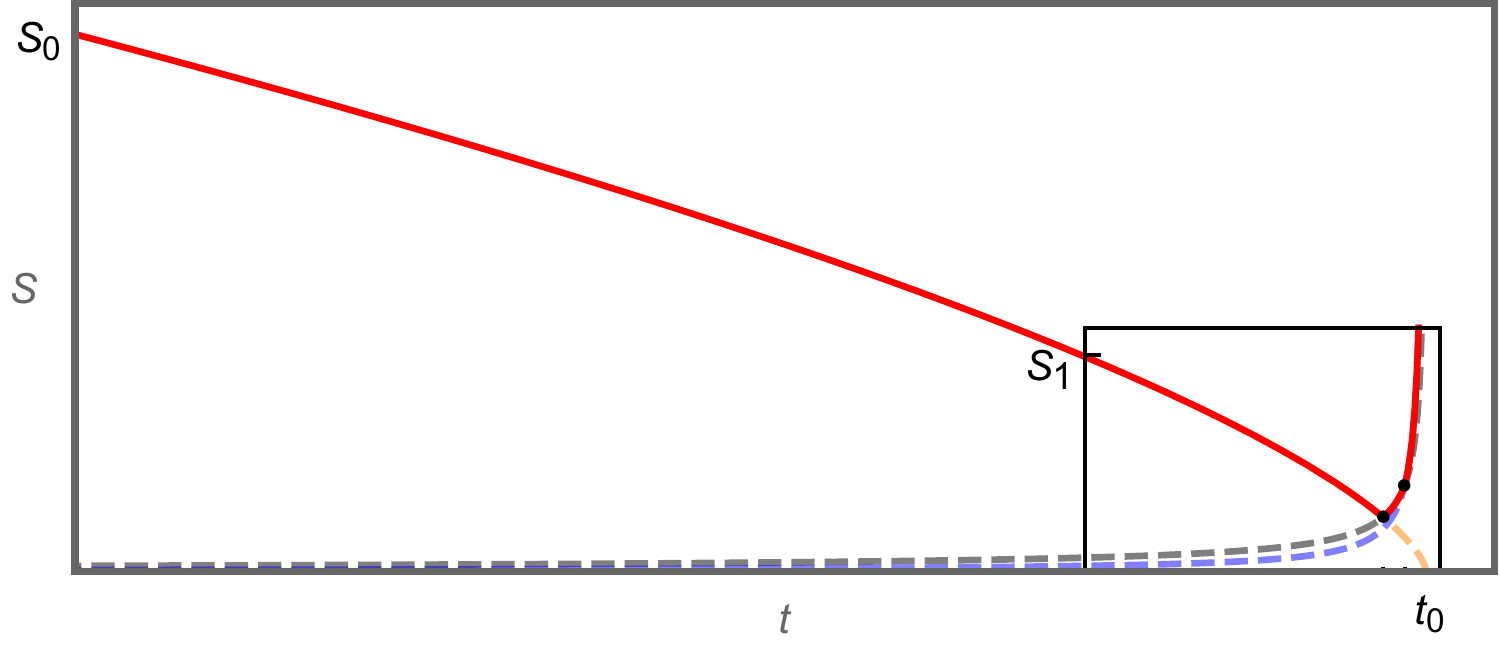}
}
\caption{
Comparison with the model with the hair-loss problem.
a) The hair-loss problem appears if the coarse-grained entropy of an evaporating black hole drops to zero; b) The entropy of the evaporating black hole increases at the end of the evaporation, and the hair-loss problem can be avoided. Note that Fig. \ref{BH} is the enlarged version of panel b) here, for the purpose of showing the transitions more clearly. The $r_s^2$ dominated regime is at early times when $r_s$ is relatively large, and the transition happens at late times.
}
\label{BH2}
\end{figure}

\subsection{Evaporating black holes}

We have analyzed the different temperature limits of U(1) gauge theory on black hole background, and see different entropy behaviors at different temperatures. Can the above result be used to understand the hair-loss problem? Let us consider an evaporating black hole whose temperature changes with the evaporation time.

First, let us look at how the Bekenstein-Hawking entropy changes with time $t$. With the Stephan-Boltzmann law, the power of the Hawking radiation is inversely proportional to the mass squared \cite{Page:2013dx}. So, the mass and horizon radius of the system can be calculated from the radiation power
\be
r_s(t)=r_0\left(1-\frac{t}{t_0}\right)^{1/3}\,,
\ee
where $r_0$ is the original horizon radius and $t_0$ is the total evaporation time.
As argued in the hair-loss problem, the coarse-grained entropy would finally decay to zero and we eventually lose all our tools for reconstructing the interior operator.
This is illustrated in Fig. \ref{BH2} (a).

However, in the new model discussed in the previous subsection, we have demonstrated a transition of entropy behaviors at high and low temperatures. 
As shown in Fig. \ref{BH}, the entropy increases at the end of the evaporation. For Schwarzschild black holes, there are two transitions as can be seen from the figure. At the beginning of the evaporation, the entropy has no difference from Page's original argument. However, after
\be
t_1=\Big(1-0.013\frac{(\sqrt{G_N}L)^{3/2}}{r_0^3}\Big)t_0\,,
\ee
the entropy starts to be proportional to $L^2/\beta^2$ as shown in \eqref{10}. As the black hole continues to evaporate, the entropy is dominated by the volume contribution shown in Eq. \eqref{7}. The time when the volume contribution overcomes all other contributions is denoted as $t_2$, as shown in Fig. \ref{BH}. $t_2$ is calculated as
\be
t_2=\Big(1-0.0096\frac{L^{3}}{r_0^3}\Big)t_0\,.
\ee
Fig. \ref{BH} is the enlarged part at late times as shown in Fig. \ref{BH2} b), for the purpose of showing the transitions more clearly. Moreover, a comparison of our model and the model with the hair-loss problem is given in Fig. \ref{BH2}.

Note that the solution for the hair-loss problem we provide here has a flavor of the black hole remnants \cite{Aharonov:1987tp, Adler:2001vs}. Generally, the end stage of evaporation is subtle and relies on a quantum theory of gravity. The advantage of our model is that the entropy behavior at the early stage of the evaporation is not changed, which means that all the conclusion from the semiclassical analysis of the information paradox is the same. But we have to admit that physics related to the end of evaporation needs further understanding, and we may gain lots of new insights by studying it.

\section{Summary}
\label{con}

In this paper, we provide a possible way of circumventing the hair-loss problem, using a gauge toy model for gravitational fluctuation on a black hole in a cavity. The hair-loss problem was put forward in studies that try to use soft hair as extra ingredients to understand the information paradox, and was demonstrated as a general problem in most schemes trying to decouple the early and late radiation to avoid AMPS firewall. The problem suggests that at the end stage of the evaporation, the small coarse-grained entropy can never suppose any tool to decouple the early and late radiation and reconstruct the interior operator. The fidelity of the reconstruction is super low. 

We study U(1) gauge theory on a black hole background with non-trivial boundary condition \eqref{BC} as a toy model to solve the problem. 
The system is put inside a cavity, and we have the black hole horizon and the surface of the cavity as the boundaries.
The boundary condition allows extra zero longitudinal momentum modes $\phi$ and boundary-anchored Wilson lines $W$. The partition function and entropy can be directly evaluated via the Euclidean path integral. How the entropy of the black hole system changes with the evaporation time is also obtained and illustrated in Fig. \ref{BH} and \ref{BH2}. There is no difference with Page's argument at the beginning of the evaporation.
However, at the end of the evaporation, rather than directly dropping to zero, the coarse-grained entropy increases, which can help us to avoid the hair-loss problem. 

Note that the strategy for circumventing the hair-loss problem here crucially relies on the specific setup in which we put the black hole inside a cavity. 
The setup largely mimics 't Hooft's brick wall model, and we have two boundaries for the gauge fluctuations on this background. 
The boundary condition \eqref{BC} is a natural one and is slightly looser than the Casimir boundary condition. The extra modes allowed by the boundary condition are the main reason that gives a large entropy at the end of the evaporation.

Recently, soft hair and other lessons learned from the BHIP were understood from regular black hole models \cite{Wang:2023tdz}, which provides a refreshing angle to understand black hole physics. Moreover, it was proposed that when non-vacuum distortions are considered, the initial data reconstruction can be realized \cite{Lochan:2016nbs, Chakraborty:2017pmn}. It is interesting to further study possible relations between the soft BHIP program and those different proposals.

\begin{acknowledgments}
We would like to thank Yang An and Pujian Mao for the useful discussions. 
This work is supported by the National Natural Science Foundation of China (NSFC) under Grant
No. 11905156, No. 11935009, and 
No. 12247101. 
\end{acknowledgments}



\providecommand{\href}[2]{#2}\begingroup\raggedright\endgroup

\end{document}